\documentclass[aps,prx,floatfix,twocolumn,nopacs,longbibliography,superscriptaddress]{revtex4-2}
\usepackage{comment}
\usepackage[utf8]{inputenc}
\usepackage[american]{babel}
\usepackage[T1]{fontenc}
\usepackage[pdftex]{graphicx}  
\usepackage{graphicx, xcolor}
\usepackage{nicefrac}
\usepackage{dcolumn}
\usepackage{bm}
\usepackage{amsmath,amsthm,amssymb}
\usepackage{amsfonts}
\usepackage{bbold}
\usepackage{hyperref}
\usepackage{xcolor}
\usepackage[normalem]{ulem}
\hypersetup{colorlinks,bookmarksopen,bookmarksnumbered,
citecolor=blue,
linkcolor=blue,
pdfstartview=false,
urlcolor=blue}
\usepackage{graphicx}
\usepackage{wrapfig}
\usepackage{soul}
\usepackage{mathtools}
\usepackage{float}
\usepackage{natbib}

\begin{document}

\global\long\def\ket#1{\left|#1\right\rangle }
\global\long\def\bra#1{\left\langle #1\right|}

\global\long\def\braket#1#2{\left\langle#1\mid#2\right\rangle }
\global\long\def\expected#1{\left\langle #1\right\rangle }

\title{Entanglement Transition due to particle losses in a monitored fermionic chain}

\author{Rafael D. Soares}
\affiliation{Université Paris-Saclay, CNRS, LPTMS, 91405 Orsay, France}
\affiliation{JEIP, UAR 3573 CNRS, Coll\`ege de France,   PSL  Research  University, 11,  place  Marcelin  Berthelot,75231 Paris Cedex 05, France}
\author{Youenn Le Gal}
\affiliation{JEIP, UAR 3573 CNRS, Coll\`ege de France,   PSL  Research  University, 11,  place  Marcelin  Berthelot,75231 Paris Cedex 05, France}
\author{Marco Schir\`o}
\affiliation{JEIP, UAR 3573 CNRS, Coll\`ege de France,   PSL  Research  University, 11,  place  Marcelin  Berthelot,75231 Paris Cedex 05, France}

\begin{abstract}
Recently, there has been interest in the dynamics of monitored quantum systems using linear jump operators related to the creation or annihilation of particles. Here, we study the dynamics of the entanglement entropy under quantum jumps that induce local particle losses in a model of free fermions with hopping and $\mathbb{Z}_2$ pairing. We solve the non-unitary dynamics using the recently developed Faber Polynomial method and explore the different steady-state entanglement regimes by tuning the pairing strength, thus interpolating between monitored free fermions 
coherently driven by a particle number conserving Hamiltonian to a parity conserving one. In the absence of pairing, all quantum trajectories approach the vacuum at long times, with the entanglement entropy showing non-monotonic behavior over time that we capture with a phenomenological quasiparticle \emph{ansatz}. In this regime, quantum jumps play a key role, and we highlight this by exactly computing their waiting-time distribution. On the other hand, the interplay between losses and pairing gives rise to quantum trajectories with entangled steady-states. We show that by tuning the system parameters, a measurement-induced entanglement transition occurs where the entanglement entropy scaling changes from logarithmic to area-law. We compare this transition with the one derived in the no-click limit and observe qualitative agreement in most of the phase diagram. Furthermore, the statistics of entanglement gain and loss are analyzed to better understand the impact of the linear jump operators.
\end{abstract}

\date{\today}
\maketitle

\section{Introduction}
There is growing interest around non-unitary noisy dynamics of many-body systems, such as those resulting from the competition between coherent evolution and stochastic quantum measurements. Out of this competition, a novel type of measurement-induced phase transition (MIPT) has been discovered in the entanglement dynamics of the system, separating a volume-law phase characteristic of unitary evolution from an area-law phase driven by the Quantum Zeno effect~\cite{li2018quantumzenoeffect,li2019measurementdrivenentanglement,skinner2019measurementinducedphase}. An intriguing aspect of this criticality is that it is encoded in the stochastic fluctuations of the measurement process and in higher-order moments of the conditional state, such as for example entanglement entropy or purity, rather than in its average state. As such, the experimental detection of MIPT is particularly challenging, even though recent progress has been made in both direct experimental evidence~\cite{noel2021measurementinducedquantum,koh2022experimentalrealizationof,hoke2023quantuminformationphases} and theoretical proposals to mitigate the so-called post-selection problem~\cite{gullans2020scalableprobesof,lu2021spacetime,li2023cross,li2023decodable,passarelli2023postselectionfree,garratt2024probing}.

On the theoretical front, MIPT has been extensively studied in the context of random quantum circuits with projective measurements~\cite{potter2022quantumsciencesandtechnology,lunt2022quantumsciencesandtechnology},
as well as using nonunitary stochastic unraveling of the Lindblad master equation for the evolution of open quantum systems~\cite{fazio2024manybodyopenquantumsystems}. In fact, these quantum trajectories, such as quantum jump dynamics~\cite{dalibard1992wavefunction,plenio1998quantum} or quantum state diffusion (QSD) protocol~\cite{gisin1992thequantumstate}, admit a natural interpretation as evolution of the system conditioned to a set of (weak)-measurement outcomes~\cite{wiseman2009quantummeasurementand}.
In the context of continuously monitored many-body systems, the primary setup typically investigated involves fermionic lattice models or quantum spin chains with local interactions and local measurements, such as those of particle density or magnetization. As for random circuits, the volume-law entanglement phase generated by unitary dynamics is expected to be robust for interacting monitored quantum systems and evidence based on finite-size numerical simulations supports this picture~\cite{fuji2020measurementinducedquantum,lunt2020measurementinducedentanglement,dogger2022generalizedquantummeasurements,xing2023interactions,altland2022dynamics}, even though the critical properties of this transition are so far unknown. On the other hand, Gaussian monitored systems are not expected to sustain a volume law in the presence of local measurements~\cite{cao2019entanglementina,chen2020emergent,fidkowski2021howdynamicalquantum}, unless the stochastic dynamics is post-selected such as, for example, in the no-click limit of the quantum jump dynamics~\cite{legal2023volumetoarea,granet2023volume}. Nevertheless, the phase diagram of non-interacting monitored systems has been the subject of several investigations, both using numerics, taking advantage of the gaussianity of the state at fixed trajectory, or within replica field theory. Monitored one-dimensional fermionic systems with strong U(1) symmetry, associated with particle number conservation, have been shown to display a crossover into an area-law phase~\cite{cao2019entanglementina,alberton2021entanglementtransitionin,vanregemortel2021entanglement,buchhold2021effective,coppola2022growthofentanglement}, with the MIPT washed away in the thermodynamic limit via a mechanism similar to weak-localization corrections~\cite{poboiko2023theoryoffree,fava2024monitored}. For fermions with discrete symmetries, such as Majorana or Ising chain, instead a genuine MIPT between sub-volume and area-law scaling of the entanglement entropy has been identified both numerically and analytically~\cite{turkeshi2021measurementinducedentanglement,botzung2021engineereddissipationinduced,bao2021symmetryenrichedphases,nahum2020entanglement,tang2021quantum,jian2022criticality,turkeshi2022entanglementtransitionsfrom,piccitto2022entanglementtransitionsin,kells2021topologicaltransitionswith,paviglianiti2023multipartite,gal2024entanglementdynamicsmonitoredsystems,fava2023nonlinearsigmamodels,jian2023measurementinducedentanglement,leung2023theoryfreefermionsdynamics}. 
\begin{figure*}[t]
    \centering
    \includegraphics{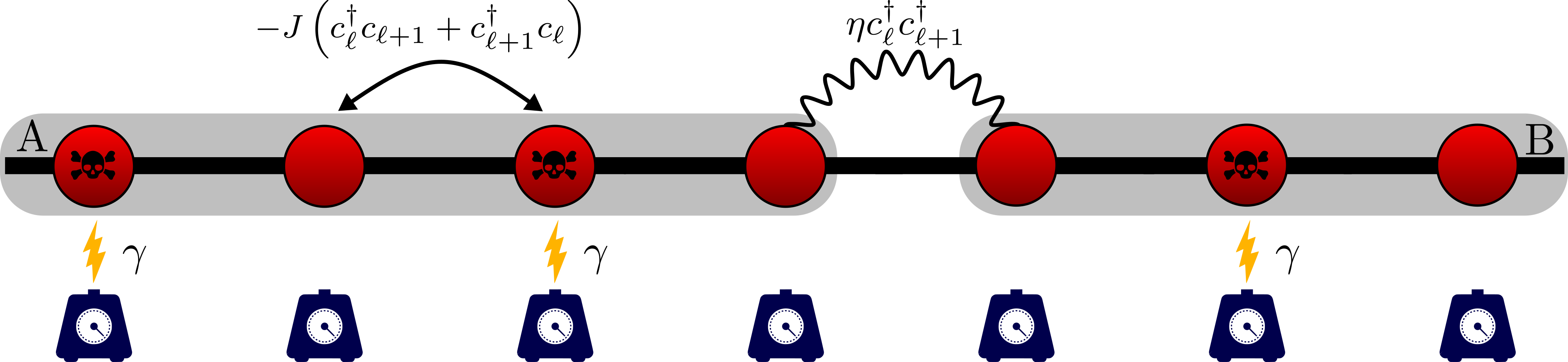}
    \caption{Scheme of the monitored fermionic chain with local particle losses happening with a rate $\gamma$, including both hopping terms controlled by $J$ and pairing terms controlled by $\eta$. At each site, there is a detector which, when triggered (represented by the lightning bolt), removes a particle from that site in the system. The A and B (the complement of A) regions represent possible partitions of the system for the computation of the entanglement entropy.}
    \label{fig:scheme_setup}
\end{figure*}

Although the role of symmetries and monitoring protocols has been discussed before, less is generally known about the role of the measurement operator on the properties of the MIPT. Recently, the entanglement dynamics of non-interacting monitored systems in the presence of jump operators which are linear, either in bosonic or fermionic operators, describing creation/destruction of quantum particles, has been considered~\cite{minoguchi2022continuousgaussianmeasurements,young2024diffusiveentanglementgrowthmonitored,yokomizo2024measurementinducedphasetransitionfree,starchl2024generalizedzenoeffectentanglement}. In these cases, the steady-state was shown to display an area-law entanglement entropy, and no MIPT was found, at least for local jump operators. However, the dynamics can be complex, displaying diffusive scaling of the entanglement entropy~\cite{young2024diffusiveentanglementgrowthmonitored} or a crossover between a critical logarithmic phase and an area-law phase~\cite{starchl2024generalizedzenoeffectentanglement}.

In this work, we consider a non-interacting fermionic chain with hopping, pairing, and monitoring through local jump operators associated with particle losses. Using the recently developed Faber Polynomial method combined with Monte Carlo wave-function~\cite{soares2024nonunitaryquantummanybodydynamics}, we show that the interplay of pairing, acting as a two-particle drive, and losses leads to a steady-state with finite density and a nontrivial entanglement structure, including a genuine MIPT. In absence of pairing, the losses lead to a depletion of the system and the entanglement entropy shows a non-monotonic behavior in time, ultimately vanishing exponentially at long times due to the effects of quantum jumps, whose waiting time distribution we compute analytically. Interestingly, we show that a simple extension of the quasiparticle picture of unitary systems is able to capture this non-monotonic effect in the entanglement entropy.
A finite pairing leads to a conditional state displaying a genuine MIPT, where the entanglement entropy in the steady-state changes from a sub-volume logarithmic scaling to an area-law. At the same time, the average state, described by a density matrix evolving under a quadratic Lindbladian, remains non-critical and characterized by a finite correlation length~\cite{zhang2022criticality}, a result that remains true even under partial averaging~\cite{paviglianiti2024breakdownmeasurementinducedphasetransitions}.

We compare the entanglement dynamics in the quantum jump trajectories with the no-click limit driven by the non-Hermitian Hamiltonian, finding an overall qualitative agreement for any finite pairing. Interestingly, we show that quantum jump dynamics leads to more entangled steady-state as compared to the no-click limit, i.e. the area-law is suppressed by quantum jumps. We interpret these findings by means of the statistics of the entanglement gain and loss~\cite{gal2024entanglementdynamicsmonitoredsystems}, a recently introduced metric that clarifies the role of quantum jumps and allows us to build a simple classical stochastic model that captures the dynamics of the entanglement under monitoring.

The paper is structured as follows. In Sec.~\ref{sec:model}, the model and the monitoring protocol are introduced. In Sec.~\ref{sec:U_1_waiting_times} we start considering the case with no pairing and compute the dynamics of entanglement entropy and the waiting-time distribution of quantum jumps. In Sec.~\ref{sec:Z2} we include a finite pairing term and provide numerical evidence for an entanglement transition between a sub-volume phase and an area-law phase as a function of the monitoring strength. We discuss the role of the pairing term in interpolating between the two entanglement scaling regimes in Sec.~\ref{sec:from_u1_to_z2}. In Sec.~\ref{sec:entanglement_statistics}, we discuss the role of quantum jumps by analyzing the statistics of entanglement gain and loss. Sec.~\ref{sec:discussion} is devoted to a discussion of our results in light of recent literature.
Finally, Sec.~\ref{sec:conclusion} contains our conclusions and future perspectives. In the Appendixes, we provide further methodological details and results relevant to our work.

\section{Model and Monitoring Protocol}\label{sec:model}
We focus our attention on the quantum jump (QJ) dynamics in a one-dimensional chain of spinless fermionic particles, see Figure~\ref{fig:scheme_setup}. The model is characterized by the following quadratic Hamiltonian
\begin{equation}
\label{eqn:H}
\begin{aligned}
\mathcal{H}=&-\sum_{n=0}^{L-2}\left[J \;c_{n}^{\dagger}c_{n+1}+\eta \;c_{n}^{\dagger}c_{n+1}^{\dagger}+\text{h.c}\right]+\\
&+h\sum_{n=0}^{L-1}c_{n}^{\dagger}c_{n},
\end{aligned}
\end{equation}
where $J$ is the hopping term, $\eta$ the pairing or driving term and $h$ an onsite potential. We are interested in the monitored dynamics under quantum jumps, where the evolution is described by the stochastic Schr\"odinger equation~\cite{ueda1990,dalibard1992wavefunction,gardiner1992wave}
\begin{align}\label{eqn:SSE}
     d\ket{\Psi(t)} &= -idt\left\{ \mathcal{H}   - \frac{i}{2}\sum_{n=0}^{L-1} \left(L_{n}^{\dagger}L_{n} - \langle L_{n}^{\dagger}L_{n}\rangle_t \right) \right\} | \Psi(t) \rangle \nonumber\\
&\ + \sum_{n=0}^{L-1} d\xi_{n}  \left\{  \frac{L_{n}}{\sqrt{\langle L_{n}^{\dagger}L_{n}\rangle}}-1 \right\} \ket{\Psi(t)},
\end{align}
where $\mathcal{H}$ is the Hamiltonian of the fermionic chain given in Eq.~\eqref{eqn:H}, $L_n,L^{\dagger}_n$ are the jump operators defined at each lattice site that describe the measurement operator, and $d\xi_n(t) \in \{0,1\}$ is an increment for the inhomogeneous Poisson process with average ${P(d\xi_{n} = 1) = dt\langle \psi(t)| L_{n}^{\dagger} L_{n} |\psi(t) \rangle}$. The QJ monitoring protocol amounts to a stochastic evolution where at random times the jump operator $L_n$ is applied to the state, followed by a normalization, while in between quantum jumps the system evolves deterministically, but under a non-unitary dynamics described by an effective non-Hermitian Hamiltonian $\mathcal{H}_{\rm eff}$ given by
\begin{equation}\label{eqn:Heff}
  \mathcal{H}_\mathrm{eff} = \mathcal{H}   - \frac{i}{2}\sum_{n} L_{n}^{\dagger}L_{n}. 
\end{equation}
We note that the evolution of the system is state-dependent (thus nonlinear); see the counter-term appearing in Eq.~\eqref{eqn:SSE}, to ensure the normalization. In this work, we consider jump operators that are linear in the fermionic operators and describe particle losses, namely
\begin{equation}\label{eqn:jump}
    L_{n} = \sqrt{2 \gamma}\; c_n,
\end{equation}
where $n$ is the lattice site and $\gamma \in \mathbb{R}^+$. For this choice of QJs the non-Hermitian Hamiltonian in Eq.~\eqref{eqn:Heff} reads
\begin{equation}\label{eqn:Heff2}
\begin{aligned}
\mathcal{H}_{\text{eff}}=&-\sum_{n=0}^{L-2}\left[Jc_{n}^{\dagger}c_{n+1}+\eta c_{n}^{\dagger}c_{n+1}^{\dagger}+\text{h.c}\right]+\\
&+\left(h-i\gamma\right)\sum_{n=0}^{L-1}c_{n}^{\dagger}c_{n},
\end{aligned}
\end{equation}
describing fermions with hopping, pairing and a complex-valued onsite potential.

The stochastic Schr\"odinger equation~\eqref{eqn:SSE} describes the evolution of the conditional state, also called the quantum trajectory. 
The average or unconditional state $\rho(t)$, obtained by taking the density matrix $\rho_{\xi}(t)=\vert\psi_{\xi}(t)\rangle
\langle\psi_{\xi}(t)\vert $ and averaging it over the stochastic noise $\xi$ in Eq.~\eqref{eqn:SSE}, i.e. $\rho(t)=\overline{\rho_{\xi}(t)}$, satisfies on the other hand the Lindblad master equation 
\begin{align}
\frac{d\rho}{dt}=-i[\mathcal{H},\rho]+\sum_n \left( L_n\rho L_n^{\dagger}-\frac{1}{2}\left\{L_n^{\dagger}L_n,\rho\right\}\right).
\end{align}
For the model of our interest, which describes free fermions with linear jump operators, this Lindblad master equation can be solved exactly, since the mixed state is Gaussian. We briefly review the properties of this solution in Appendix~\ref{sec:lindblad}.

The descriptions in terms of quantum trajectories for pure states or Lindblad evolution for the density matrix are equivalent for what concerns physical observables, which are linear functionals of the conditional state, and indeed quantum trajectories are also called \emph{unravelings} of the Lindblad master equation~\cite{carmichael1999statisticalmethodsin}. However, the full stochastic dynamics gives access to the random fluctuations of the monitoring protocol. As often in disordered systems, whenever one is interested in quantities which are not self-averaging, these fluctuations play a crucial role. This is the case of higher moments of the density matrix, relevant, for example, to compute the purity of the state, its entanglement entropy, connected correlation functions or overlaps~\cite{turkeshi2021measurementinducedentanglement}. These contain physics that is not captured by the averaged state. The relevant quantity considered is the von Neumann entanglement entropy, defined as ~\cite{calabrese2004entanglemententropyand,amico2008entanglementinmanybody}
\begin{equation}
S_\mathbf{\xi}(t,\xi_t )  = -\mathrm{tr}_A\left[ \rho_\mathbf{\xi}^A(t)\ln \rho_\mathbf{\xi}^A(t)\right]\;,
	\label{eq:5v0}
\end{equation}
where we have introduced a bipartition $A\cup B$ in the system (cf. Fig.~\ref{fig:scheme_setup}), with the reduced density matrix $\rho_\mathbf{\xi}^A(t) = \mathrm{tr}_B |\Psi_\mathbf{\xi}(t)\rangle\langle\Psi_\mathbf{\xi}(t)|$. In the following, we are interested in the average entanglement entropy, given by 
\begin{equation}\label{eqn:aver_s}
    \overline{S(t)} = \int \mathcal{D}\mathbf{\xi}\; P(\mathbf{\xi}) S_\mathbf{\xi}(t),
\end{equation}
where the average is taken over the Poisson measurement noise $\xi$. In particular, we will discuss how the steady-state entanglement entropy scales with the system's size and the possibility of entanglement phase transitions driven by the monitoring rate.

It is useful to briefly discuss the symmetries of our problem. We first note that in the absence of pairing $\left(\eta=0\right)$, the Hamiltonian has U(1) invariance 
$\left(c_n\rightarrow e^{i\phi}c_n, \mbox{for an arbitrary phase}\,\phi \right)$.
In a closed system this symmetry would imply the conservation of the total particle number; however, the conservation is broken in the stochastic QJ dynamics because of the jump operator associated with particle losses. Moreover, at the level of the average state, described by the Lindblad master equation,  the system exhibits weak U(1) symmetry. As is typical for open quantum systems, such a symmetry does not imply the existence of conserved quantities~\cite{fazio2024manybodyopenquantumsystems}. Specifically, for $\eta=0$, the number of particles is not conserved due to losses even in the Lindblad dynamics.

For $\eta\neq0$ the pairing term in $\mathcal{H}$ reduces the U(1) symmetry down to a $\mathbb{Z}_2$ one, associated with the fermionic parity $\left(c_n\rightarrow -c_n\right)$. In this case, the particle number is not conserved in the coherent dynamics. In the Lindblad dynamics, this corresponds to a weak $\mathbb{Z}_2$ symmetry, as the parity changes over time as a result of particle losses. At the level of quantum trajectories, if one would start from an eigenstate of the fermionic parity, the non-Hermitian Hamiltonian would preserve it during the evolution, whereas quantum jumps alter the parity of the state. This results in the fermion number being switched between the even and odd sectors (or vice versa) after each quantum jump. However, since quantum jumps act instantaneously, this does not result in a coherent superpositions of states with different fermionic parities. Consequently, the expectation value of the fermionic operators $\langle  c_n \rangle $ is zero at all times in the QJ protocol, as it should be. In this respect, the situation would be different for the quantum state diffusion protocol~\cite{gisin1992thequantumstate,landi2023current}, which in the present case would require a continuous monitoring of the expectation value of the fermionic creation/annihilation operator, which is unphysical. As such, we emphasize that for quantum state diffusion linear jump operators are physical only for bosons but not for fermions.

\begin{figure*}[t] 
\includegraphics{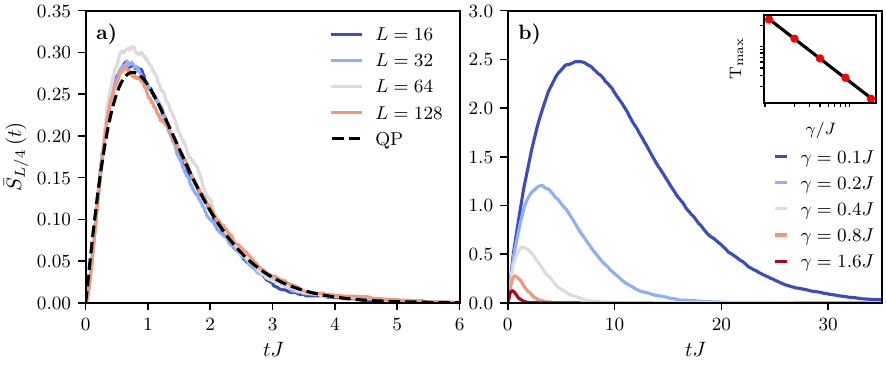}
\caption{\label{fig:U1_entanglement_entropy}  
    Average dynamics of the entanglement entropy for a cut $L/4$, under the monitored free fermion chain with $\eta=0$ and particle losses. In the left panel ($a)$) we fix $\gamma=0.8J$ and show that results are independent on system size and compare well with the quasiparticle picture (QP) given by Eq.~\ref{eq:qp_pic}. In the right panel ($b)$), the dependence of the entanglement entropy with $\gamma/J$ for a system with fixed size $L=128$ is shown. The inset on the right panel shows the time $\text{T}_{\max}$ for the entanglement entropy to reach its maximum. This time scales with $\sim\gamma^{-\alpha}$, where $\alpha=\left(1.0\pm0.1\right)$. In both panels, the system is prepared at half-filling. Other parameters: $h=0$ and $\eta=0$.}
\end{figure*}

Our model arises naturally in the context of dissipative extensions of the Kitaev chain~\cite{Kitaev_2001}, where it describes electrons with p-wave pairing of amplitude $\eta$, and additionally particle losses. Its Lindblad dynamics has been studied before in various works (see for example Refs.~\onlinecite{eisert2010noisedrivenquantumcriticality,diehl2011topology,moos2019dynamical,lieu2020tenfold,starchl2024quantum}), while the entanglement dynamics of quantum trajectories has not been explored before to the best of our knowledge. We also emphasize that our model is different from a dissipative or monitored quantum spin chain of Ising or XY type, studied, for example, in Refs.~\onlinecite{prosen2008quantum,joshi2013quantum,kavanagh2022effects}, due to our choice of jump operators in Eq.~\eqref{eqn:jump} that does not include the Jordan-Wigner string~\cite{mbeng2024thequantum}, thus breaking the equivalence upon fermionization of the spin. However, in the no-click limit, where quantum jumps are neglected, our non-Hermitian Hamiltonian in Eq.~\eqref{eqn:Heff2} reduces to the one obtained for a monitored Ising chain, whose entanglement dynamics and the steady-state properties have been studied in detail~\cite{turkeshi2023entanglementandcorrelation,zerba2023measurement}.

Before concluding this section, we briefly mention how we solve the dynamics of the QJ trajectories. Since the Hamiltonian we consider is quadratic and the jump operators linear, the system remains in a fermionic Gaussian state along a quantum jump trajectory~\cite{bravyi2005lagrangian,landi2023current}. Therefore, all the information on the state is encoded in the 2-point correlation matrix. To solve the quantum jump dynamics of our model, we resort to the recently developed Faber Polynomial method~\cite{soares2024nonunitaryquantummanybodydynamics}. Specifically, we solve the non-Hermitian dynamics between quantum jumps by expanding the non-unitary evolution operator in a single-particle basis in Faber polynomials. To apply the quantum jump, we compute the correlation matrix and update it after the jump. Finally, from the dynamics of the correlation matrix, we can directly obtain the entanglement entropy. The details of the numerical methods used are discussed in Appendix~\ref{sec:details}. Throughout this work, we consider a system with open boundary conditions and an initial state that is a product state of the form $\ket{\Psi_0} =\prod_{n=0}^{L/2} c^\dagger_{2n} \ket{\text{vac}}$.

\section{Monitoring Dynamics in Absence of Pairing}
\label{sec:U_1_waiting_times}
We start by discussing the QJ dynamics in the case $\eta=0$, corresponding to free fermions with losses and without pairing. In this case, the evolution in-bewteen quantum jumps conserves the total particle number. However, the quantum jump operators break this conservation by destroying a fermion, and so, we expect the system, starting with a well-defined number of particles, to approach the vacuum for sufficiently long times under stochastic dynamics.  We verify this both at the level of the average state (Lindbladian dynamics) (see Appendix~\ref{sec:lindblad}) and at the level of the unraveled quantum trajectory.

In Fig.~\ref{fig:U1_entanglement_entropy} $a)$, we plot the time evolution of the average entanglement entropy for a cut of size $\ell=L/4$ and different system sizes. We see that the entropy at short time grows linearly in time, as it would under the unitary evolution, then reaches a maximum and decreases towards zero. The successive application of the jump operator removes the particles and drives the entanglement entropy towards zero, as we are left with the fermionic vacuum, a product state. The dynamics of entanglement entropy is essentially independent of $L$, for initial states with the same filling. This should be the case, as the probability of jumping after a time, $dt$, at any site is given by $p_{\rm jump} = \gamma dt N_0$ so the probability of jumping per total lattice size is independent of the total system size.

In Fig.~\ref{fig:U1_entanglement_entropy} $b)$, we plot the dependence of the entanglement dynamics on the monitoring rate $\gamma$. We see that the latter controls (i) the time at which the monitoring sets in and the entanglement entropy starts to decrease and (ii) the value of the maximum entanglement entropy that can be reached. As we see in the inset, the time scales for the crossover is essentially $\text{T}_{\rm max}\sim 1/\gamma$, which also corresponds to the value of the maximum entanglement entropy. The interpretation of this scale is clear from the point of view of the quasiparticle picture for the entanglement dynamics~\cite{Calabrese_2005,Calabrese_2016,Calabrese_2020_lecture_notes,cao2019entanglementina,turkeshi2022entanglementtransitionsfrom}: 
in the unitary case the entanglement entropy is carried by pairs of quasiparticles propagating ballistically across the cut, leading to a linear growth in time, while
monitoring gives a characteristic resetting time scale to the quasiparticles, which stops the growth of entanglement to a value of order $1/\gamma$. 
This qualitative argument explains the deviation from unitary dynamics, but it does not account for the long-time behavior when the system depletes and the entropy vanishes. 

The full-time evolution of the entanglement entropy can be understood more quantitatively using a simple extension of the quasiparticle picture.
We note that for the average mixed state described by a quadratic Lindblad master equation, the quasiparticle picture was developed in Ref.~\cite{carollo2022dissipative}. Here, instead, we focus on the conditional state of the monitored system, i.e. on the entanglement of the quantum trajectory. Specifically, we assume that due to quantum jumps the quasiparticle occupation number $n_k=c^{\dagger}_kc_k$ (which in the present case of $\eta=0$ corresponds to the particle occupation number) is effectively decreasing exponentially in time, according to the expression
\begin{equation}
    \left \langle n_k \left(t \right) \right\rangle  = \left \langle n_k \left(0 \right) \right\rangle e^{-\alpha t}, 
\end{equation}
where $\alpha \in \mathbb{R}_+$ is an effective decaying rate. This is the case at the level of the Linbladian dynamics, where $\alpha = 2\gamma$ (see Appendix~\ref{sec:lindblad}). Using this expression directly in the quasiparticle formula~\cite{Calabrese_2005,Calabrese_2016,Calabrese_2020_lecture_notes,young2024diffusiveentanglementgrowthmonitored}, the entanglement entropy is given as
\begin{equation}
    \bar{S}_{L/4} \left(t \right) \propto \int_{-\pi}^{\pi} \dfrac{dk}{2\pi} \min \left(2 t \left| v_k \right|, L / 4 \right) H\left(\left \langle n_k \left(t \right) \right\rangle\right),
    \label{eq:qp_pic}
\end{equation}
where $v_k=-2J\sin(k)$ represents the semi-classical velocity determined by the Hermitian part of the Hamiltonian, given that the non-Hermitian term merely adds an overall background decay, and $H(x)=-x\ln x - (1-x)\ln\left(1 - x \right)$. This expression matches the numerical results, yielding a renormalized $\alpha$ coefficient, in contrast to the Linblad case. Nevertheless, $\alpha \propto \gamma$ as shown in Appendix~\ref{app:quasiparticle}. This corroborates the dependence in $\text{T}_{\max}$ with $\sim 1/\gamma$.

\begin{figure*}[t] 
\centering
\includegraphics{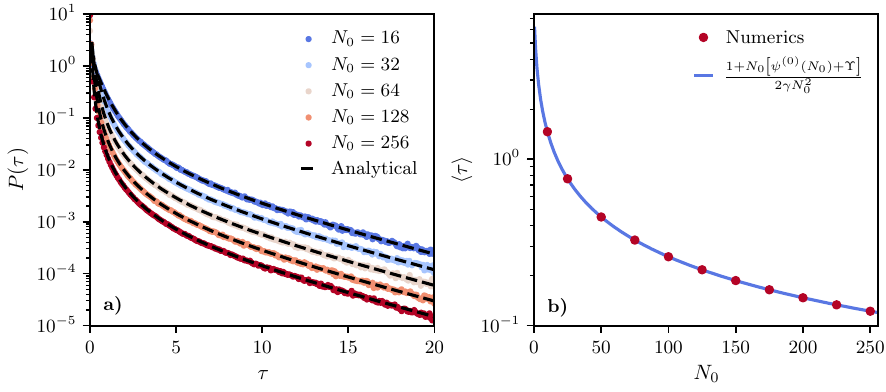}
\caption{\label{fig:wtd_U1_case} Statistics of QJs Waiting Times for monitored fermions in the absence of pairing. $a)$ - Waiting Time Distribution for different values of $N_0$. $b)$ - Average waiting time as a function of $N_0$ for $\eta=0$. Other parameters: $\gamma = 0.1J$.} 
\end{figure*}

It is interesting to compare the QJ dynamics with the result of the non-Hermitian evolution corresponding to the no-click limit. For $\eta=0$ the non-Hermitian Hamiltonian in Eq.~\eqref{eqn:Heff2} simplifies, since its real and imaginary parts commute among each other, i.e. $\mathcal{H}_{\rm eff}=\mathcal{H}-i\gamma N$ with $N$ the total particle number operator and $\mathcal{H}=-\sum_{n}J c^{\dagger}_nc_{n+1} +\mbox{h.c.}+h N$, with $[\mathcal{H},N]=0$. As a consequence, if one starts with an initial state $\vert\Psi(0)\rangle$ which is an eigenstate of the total particle number with eigenvalue $N_0$ the evolution of the wave function takes the form
\begin{equation}
\ket{\Psi\left(t\right)}  = \dfrac{e^{-\gamma t\cdot N_0} e^{-it\mathcal{H}}\ket{\Psi\left(0\right)}}{e^{-\gamma t\cdot N_0} \left\Vert e^{-it\mathcal{H}}\ket{\Psi\left(0\right)}\right\Vert}\,.
\end{equation}
From this expression we see that the non-Hermitian term only gives rise to a trivial exponential decay that cancels out with the normalization, leaving an evolution which is completely unitary and driven only by the real part of $\mathcal{H}_{\rm eff}$, describing fermions hopping on the lattice. Furthermore, we see that the number of particles is conserved in this case.

So, in this case, the no-click limit corresponds to the unitary evolution. Quasi-particles propagate through the system with constant velocity leading to an initial linear growth in time of the entanglement entropy and saturating, at late times, to a quantity that is extensive in the subsystem size, i.e. a volume-law scaling, as the system effectively thermalizes according to the appropriate (generalized) Gibbs ensemble. As such, there is no entanglement transition associated with the non-unitary dynamics, as the system is insensitive to the non-Hermitian term.  We note however that if the initial state is not an eigenstate of the total particle number, the system will relax to a state within the sector containing the lowest particle number, as this corresponds to the slowest decaying mode. That is, the vacuum whenever $\braket{\text{vac}}{\Psi (t_0)} \neq 0$.  To conclude, in this case the QJ dynamics differs qualitatively from the no-click limit, and jumps are a relevant perturbation.

\subsection{Waiting Time Distribution}

To better understand the role of quantum jumps, we examine the waiting time distribution (WTD) of successive quantum jumps~\cite{albert2012electron,landi2023current}. In this specific model, it is possible to derive an exact analytical expression for this quantity when the initial state is an eigenstate of the total particle number.

First, one should recall that for a given number of particles $N$, or equivalently for a stochastic dynamics with strong U(1) symmetry, such as particle density monitoring~\cite{gal2024entanglementdynamicsmonitoredsystems}, the waiting time distribution is Poissonian,
\begin{equation}\label{eqn:P_pois}
\mathcal{P}\left(\tau,N\right)=2\gamma Ne^{-2\gamma N\tau},
\end{equation}
with an average waiting time $\langle \tau\rangle\sim 1/\gamma N$. Given a fixed filling $n=N/L$, this implies a vanishing average waiting time in the thermodynamic limit. In our case, we should interpret Eq.~\eqref{eqn:P_pois} as a joint probability distribution, since the particle number varies stochastically in integer steps of $1$ from its initial value, $N_0$, to zero. The correspondent WTD is obtained by performing the sum $\mathcal{P} \left(\tau\right) = \sum_{N=0}^{N_{0}} \mathcal{P} \left(\tau, N \right) $. Using the usual trick of performing the sum via the primitive of $\mathcal{P}\left(\tau,N \right)$ and then deriving with respect to $\gamma$, one obtains the WTD in a closed form,
\begin{equation}\label{eqn:wtd}
\mathcal{P}\left(\tau\right) =\dfrac{\gamma}{N_{0}}\frac{1-e^{-2\gamma N_{0}\tau}\left(1+N_{0}\left(1-e^{-2\gamma\tau}\right)\right)}{2\sinh^{2}\left(\gamma\tau\right)}.
\end{equation}
In Fig.~\ref{fig:wtd_U1_case} $a)$, we sample the WTD through our Monte Carlo wave function algorithm and compare it with the analytical prediction of Eq.~\eqref{eqn:wtd} for different $\gamma$, showing a perfect match. We see that the behavior at short time is controlled by the initial particle number, while the tails of the distribution controlling the largest waiting times are set by the states with the smallest particle number, namely when the number of particles is one. This can be read from the pre-factor of the exponential,
\begin{equation}
\lim_{\tau\rightarrow+\infty}\mathcal{P}\left(\tau\right)=\dfrac{2\gamma}{N_{0}}e^{-2\gamma\tau}.
\end{equation}
As we have access to the full WTD, we can also extract other quantities such as the average waiting time,

\begin{figure*}[t]
\centering
\includegraphics{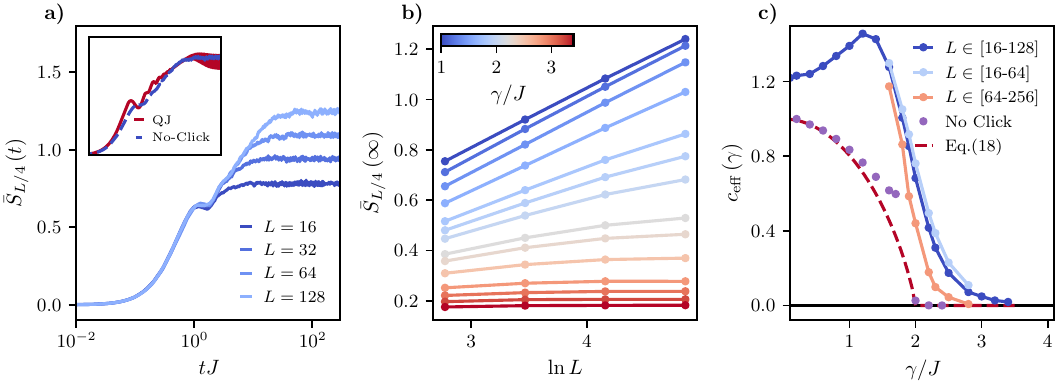}
\caption{$a)$ - Entanglement Entropy Dynamics in the Monitored $\mathbb{Z}_2$ Chain for different total system sizes with $\gamma = 0.8J$. The inset compares the dynamics in the no-click limit with that involving quantum jumps for $L=128$. $b)$ - Entanglement entropy as a function of the total size of the system for different strengths of the monitoring rate. For the values of $\gamma =\left[1.6-2.8\right]$, the fit was performed using also results for $L=256$. $c)$ - Effective central charge as a function of the monitoring parameter. The red dashed curves correspond to the calculations of the central charge in the thermodynamic limit given by Eq.~\eqref{eq:ceff_thermody}, while the remaining curves correspond to the central charge, obtained from dynamics with quantum jumps, extracted by performing the fit with different total system sizes as indicated in the legend. The purple dots were obtained through a finite-sized scaling analysis of the no-click dynamics. Other parameters: $\eta=J$ and $h=0$.}
\label{fig:Ising_line}
\end{figure*}

\begin{equation}
\left\langle \tau\right\rangle =\frac{1+N_{0}\left[\psi^{(0)}\left(N_{0}\right)+\Upsilon\right]}{2\gamma N_{0}^{2}},
\end{equation}
where $\psi^{(0)}\left(x\right)$ is the zero order poly-gamma function
and $\Upsilon=\int_{1}^{\infty}\left(-\frac{1}{x}+\frac{1}{\left\lfloor x\right\rfloor }\right)dx$
is the Euler-gamma constant. The average waiting time is controlled by the timescale set by the measurement rate, $\gamma$, as $\langle\tau\rangle\propto\gamma^{-1}$. In the limiting
case, where $N_{0}\gg1$, the average waiting time goes as,
\begin{equation}
\left\langle \tau\right\rangle \simeq \dfrac{1}{2\gamma} \left[ \frac{\log(N_{0})}{N_{0}}+\dfrac{\Upsilon}{N_{0}}+\frac{1}{2 N_{0}^{2}}\right]+\mathcal{O}\left(N_{0}^{-3}\right).
\end{equation}
These results perfectly match the numerics, as shown in Fig.~\ref{fig:wtd_U1_case} $b)$.

\section{Monitoring Dynamics with Pairing}\label{sec:Z2}

In this Section, we discuss the dynamics of the entanglement entropy for the full model in Eq.~\eqref{eqn:H}-\eqref{eqn:SSE}. Initially, we set $\eta=J$ and $h=0$ and examine how the dynamics vary with the ratio of the monitoring rate to the hopping, $\gamma/J$. Subsequently, we explore the influence of the onsite energy $h$ and pairing amplitude $\eta$. 

\subsection{Measurement-Induced Entanglement transition}

Contrary to the case discussed before (with $\eta=0$),  a finite pairing amplitude $\eta$ breaks the conservation of particles number in the coherent dynamics and act as drive mechanism that counterbalance the losses induced by the quantum jumps. As a result the pairing term in the Hamiltonian allows quantum trajectories to reach a non-equilibrium steady-state with a finite density of particles (see Appendix~\ref{sec:lindblad}) and, as we are going to discuss, achieve a non-trivial entanglement structure. 
In Fig.~\ref{fig:Ising_line} $a)$, we plot the time evolution of the average entanglement entropy for a cut $\ell=L/4$ and various system sizes with $\gamma=0.8J$. The entanglement entropy is observed to increase over time and after a short-time regime independent of the system size, we observe a logarithmic growth in time which then approaches a steady-state value which depends on $L$.

It is interesting at this point to compare the stochastic dynamics with the no click evolution, driven by $\mathcal{H}_{\rm eff}$ in Eq.~\eqref{eqn:Heff}. This is shown in the inset of Fig.~\ref{fig:Ising_line} $a)$. The time evolution of the entanglement entropy is strikingly similar in the two cases, with a logarithmic growth in time and a saturation to a steady-state value also scaling as the logarithm of the subsystem size. These numerical results for the no-click limit are fully consistent with the exact computations of the entanglement entropy in the thermodynamic limit~\cite{turkeshi2023entanglementandcorrelation}. The similarity between stochastic dynamics and the no-click limit in the small monitoring regime was also observed for the Ising chain with particle density monitoring under the QJ and QSD protocols~\cite{turkeshi2021measurementinducedentanglement,gal2024entanglementdynamicsmonitoredsystems}. This result is particularly remarkable because in this regime the average waiting-time between quantum jumps is of the order $1/L$, as we have verified numerically, which means there are many jumps along a typical quantum trajectory, yet the entanglement entropy is well captured by the no-click limit. We will come back to this point later in Sec.~\ref{sec:entanglement_statistics}.

We now study more systematically the steady-state entanglement entropy scaling by computing it numerically for different total system sizes, $L$, for a subsystem of fixed sub-size $\ell/L=\nicefrac{1}{4}$ as shown in Fig.~\ref{fig:Ising_line} $a)$. From this we see that increasing the monitoring rate $\gamma$ drives a MIPT from a sub-volume critical phase with logarithmic scaling of the entanglement entropy ($S\propto\ln(L)$) to an area-law phase ($S\propto\mathcal{O}(1)$).

\begin{figure*}
\begin{centering}
\includegraphics{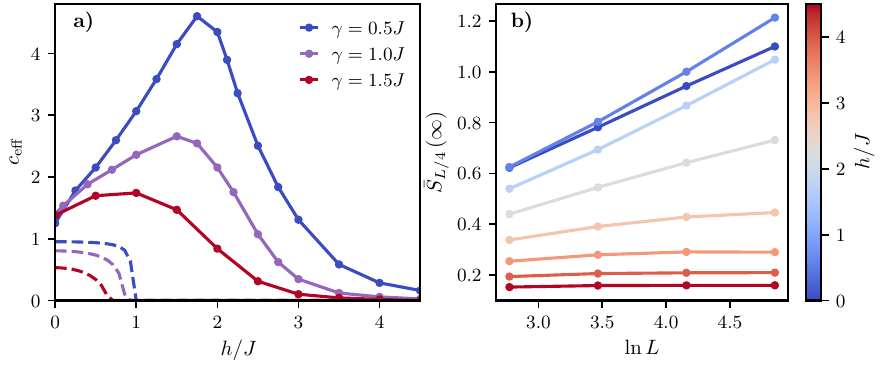}
\end{centering}
\caption{$a)$ - 
 Effective central charge as a function of the onsite potential for different values of $\gamma$. The dashed curves correspond to the No-Click limit computed from Eq.~\eqref{eq:ceff_thermody}. $b)$ - Steady state average entanglement entropy as a function of the total size of the system for different strengths of the local potential $h/J$ with the monitoring rate $\gamma=1.5J$.
Other parameters: $\eta=J$\label{fig:c_eff_onsite_potential}.
}
\end{figure*}

To quantify this transition, we proceed phenomenologically and use the formula for entanglement entropy of a Conformal Field Theory~\cite{cardy2004boundary,Calabrese_2009},
\begin{equation}\label{eqn:scft}
S=\dfrac{c_{\rm eff}}{6}\ln\left(\dfrac{2L}{\pi}\sin\left(\dfrac{\pi\ell}{L}\right)\right)+s_0,
\end{equation}
to extract an effective central charge $c_{\rm eff}(\gamma)$ which depends on the monitoring rate $\gamma$. In the above expression, $L$ the total system size and $s_0$ a non-universal constant term of order $\mathcal{O} \left(1\right)$. 

At this point, it is instructive to recall the results for the entanglement entropy in the purely non-Hermitian no-click case, where exact calculations can be done in the thermodynamic limit~\cite{turkeshi2023entanglementandcorrelation}. Interestingly, one finds a scaling behavior similar to Eq.~\eqref{eqn:scft} in the weak monitoring phase with an effective central charge $c_{\rm eff}$  which can be written in closed form as
\begin{equation}
c^{\rm no-click}_{\rm eff}=\dfrac{12}{\pi^{2}}\text{Re}\int_{0}^{1}f(\lambda)\dfrac{\lambda}{1-\lambda^{2}}\dfrac{\sqrt{1-\beta^{2}}}{\sqrt{\lambda^{2}-\beta^{2}}},
\label{eq:ceff_thermody}
\end{equation}
where $f\left(x \right) = - \frac{1-x}{2}\ln\left(\frac{1-x}{2} \right) - \frac{1+x}{2}\ln\left(\frac{1+x}{2} \right)$ and $\beta = \gamma/\left(2\eta\sqrt{1-h^2}\right)$. The system undergoes therefore an entanglement transition from a critical phase with logarithmic scaling of the entanglement entropy to an area-law scaling for values of $\gamma$ given by
\begin{equation}
    \gamma = \dfrac{2\eta}{\sqrt{1-h^2}}.  
\end{equation}
As reported~\cite{turkeshi2023entanglementandcorrelation}, this entanglement transition is controlled by a spectral transition from gapless to gapped in the imaginary part of the spectrum of the non-Hermitian effective Hamiltonian.

In the following, we use this result to compare with the QJ case. In Fig.~\ref{fig:Ising_line} $c)$, we plot the behavior of the effective central charge as a function of $\gamma/J$ for both QJs and no-click evolution. In both cases, we see that $c_{\rm eff}$ vanishes at a critical monitoring rate which signals the entanglement transition into the area-law. In the no-click limit, the numerical estimate for $c_{\rm eff}$ matches well the exact result in the thermodynamic limit~\cite{turkeshi2023entanglementandcorrelation}, which suggests that finite-size effects are limited, except close to the transition point. In the QJ case, we see that in the weak monitoring regime $c_{\rm eff}$ has a non-monotonic behavior, growing and then decreasing, and that at large $\gamma$ the effective central charge vanishes. Interestingly, we find that the effective central charge obtained in the QJ protocol is higher than that of the no-click limit; that is, jumps can lead to an increase of the entanglement entropy, an effect observed also for particle density monitoring~\cite{gal2024entanglementdynamicsmonitoredsystems}. To summarize, our results for the entanglement entropy show that an entanglement transition driven by particle losses emerges both in the QJ trajectories and in the no-click limit. In spite of the strong similarity between the two dynamics, there are also differences: in the no-click limit, the transition was associated to the opening of a gap in the imaginary part of the quasiparticle spectrum at $\gamma/J=2$. In the QJ case, the quantum jumps renormalize the critical value of $\gamma$ where this transition takes place. 

\begin{figure*}[t!] 
\centering
\includegraphics{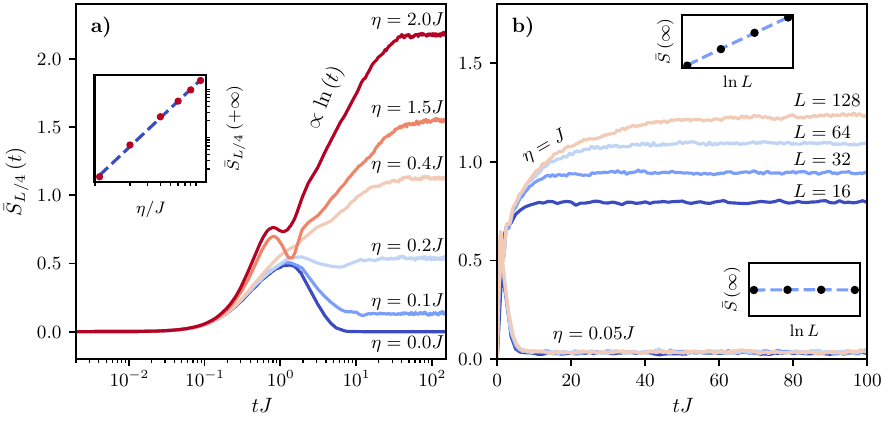}
\caption{\label{fig:U1_to_z2} Entanglement dynamics for different pairing strengths.  $a)$ - Time evolution for the entanglement entropy for different values of the pairing term. For small values of $\eta/J$ the steady-state entanglement entropy grows proportionally to $\eta^\alpha$, with $\alpha=\left(1.98 \pm 0.03 \right)$. $b)$ - Time evolution for different system sizes with $\eta = J$ (log phase) and $\eta=0.05J$ (area phase). Other parameters: $L=128$, $\gamma = 0.5J$ and $h=0$.}
\end{figure*}

\subsection{Role of the Onsite Potential \texorpdfstring{$h$}{h}}

In the results presented until now, we have always assumed that the onsite potential $h$ was zero. This parameter does not affect the dynamics in absence of pairing, as each quantum trajectory remains an eigenstate of the total particle number operator, so this term contributes only to a global phase shift of the wavefunction. However, in presence of a finite pairing $\eta$ when the coherent Hamiltonian has $\mathbb{Z}_2$ symmetry, this is no longer true. In fact in the no-click limit, the log-to-area entanglement transition can also be induced by tuning $h$ while keeping $\gamma/J$ and $\eta/J$ fixed~\cite{turkeshi2023entanglementandcorrelation}. This occurs once more due to the spectral transition from a gapless imaginary phase to a gapped one, in the no-click Hamiltonian.

In the case of quantum jumps we repeat the analysis of the scaling of steady-state entanglement entropy and extract the effective central charge $c_{\rm eff}$ as discussed in the previous section. In Fig.~\ref{fig:c_eff_onsite_potential} a), we plot its dependence with $h/J$ for different values of $\gamma/J$ and compare again with the expression in the no-click limit, Eq.~\eqref{eq:ceff_thermody}. For completeness, we also plot in Fig.~\ref{fig:c_eff_onsite_potential} $b)$ the data for the entanglement entropy corresponding to a $\gamma=1.5J$ from which the corresponding effective central charge is extracted.
As for the no-click case, we also see that in the full stochastic case for increasing values of $h/J$ the steady-state entanglement entropy undergoes a transition from a sub-volume logarithmic phase into an area-law phase. The dependence with $h$ is non-monotonous, with a maximum that is particularly pronounced for small $\gamma$. In the quantum jump case, the transition is pushed to higher values of $h$, compared to the no-click limit, an effect that was also observed for the Ising chain with particle density monitoring~\cite{paviglianiti2023enhanced, gal2024entanglementdynamicsmonitoredsystems}. Upon increasing the monitoring rate $\gamma$, the critical value of $h$ for the transition decreases, i.e. the area-law is promoted, which confirms the qualitative shape of the phase diagram. Interestingly for intermediate values of $h/J$ and weak monitoring the logarithmic phase is enhanced, i.e. quantum jumps can lead to a more entangled steady-state than the no-click evolution. We will comment on this feature in the next section.

\subsection{Role of the Pairing Amplitude}
\label{sec:from_u1_to_z2}

From the previous two sections, we conclude that the entanglement dynamics in the absence of pairing, where the coherent dynamics exhibits U(1) symmetry and so the particle losses drive the system towards the vacuum, is substantially different from the case with finite pairing, where the $\mathbb{Z}_2$ symmetry of the coherent dynamics leads to a non-trivial steady-state under quantum jumps.

We now investigate how these two limits are connected by studying the QJ dynamics as a function of the pairing amplitude $\eta$ and the monitoring rate $\gamma$. In Fig.~\ref{fig:U1_to_z2} $a)$, the dynamics of the entanglement entropy is shown for increasing values of $\eta$. At short times, the entanglement entropy increases and then halts after a timescale determined by $\sim \gamma^{-1}$. However, a finite pairing $\eta$ counterbalances the particle losses, leading to a finite entanglement in the steady-state, growing as $(\eta/J)^2$ for $\eta \ll \gamma$(see inset). As $\eta/J$ increases further, the entanglement entropy does not decrease after this initial period; instead, the finite pairing term allows for a logarithmic ($\ln t$) entanglement production. In Fig.~\ref{fig:U1_to_z2} $b)$, we depict the behavior of entanglement for two different values of $\eta/J$ across various system's sizes. When the monitoring rate is less dominant compared to the pairing term, the steady-state entanglement entropy exhibits sub-volume scaling consistent with a logarithmic law. Conversely, if the monitoring rate prevails over the pairing term, the steady-state entanglement entropy scales according to an area-law. The findings indicate that a phase transition in the steady-state entanglement of QJ dynamics can also be triggered by $\eta/J$. In other words, MIPT occurs at a critical point that is significantly influenced by $\eta/J$. In general, reducing the pairing term narrows the sub-volume log-law phase, promoting an area-law scaling for the entanglement entropy, again in qualitative agreement with the no-click results~\cite{turkeshi2023entanglementandcorrelation}.

\section{Statistics of Entanglement Gain and Loss}
\label{sec:entanglement_statistics} 

In this Section, we analyze in detail the role of quantum jumps on the entanglement dynamics of our monitored chain by resorting to a recently introduced metric~\cite{gal2024entanglementdynamicsmonitoredsystems}: the statistics of the entanglement gain and loss. 

We begin with a brief overview of this concept: A quantum trajectory is composed of non-Hermitian evolution interspersed with random QJs. The entanglement entropy evolves smoothly under the non-Hermitian evolution, whereas the QJs lead to discontinuous changes. The key idea of this approach is to sample the specific variations in the entanglement entropy caused by each QJ, conditioned to the \emph{entanglement content} $S$, i.e. the state's entanglement entropy at a specific time before the random event. This leads to a probability distribution $P(\Delta S_\mathrm{qj}\lvert S)$. Similarly, for the non-Hermitian evolution between QJs, instead of looking at the absolute change over the waiting time $\Delta t$, we focus on the rate $(S_{t+\Delta t } - S_{t})/(\Delta t )$, as it is not an instantaneous event like a QJ. This approach results in the distribution $P(\delta S_\mathrm{nH}\lvert S)$. As shown in~\cite{gal2024entanglementdynamicsmonitoredsystems}, these probability distributions play a key role in the dynamics of the entanglement entropy. On the one hand, they encode the mutual impact of quantum jumps and non-Hermitian evolution on the entanglement entropy. On the other hand, they allow us to build an effective classical stochastic model for the entanglement dynamics, as we will discuss later on.

\begin{figure*}[t!] 
\centering
\includegraphics {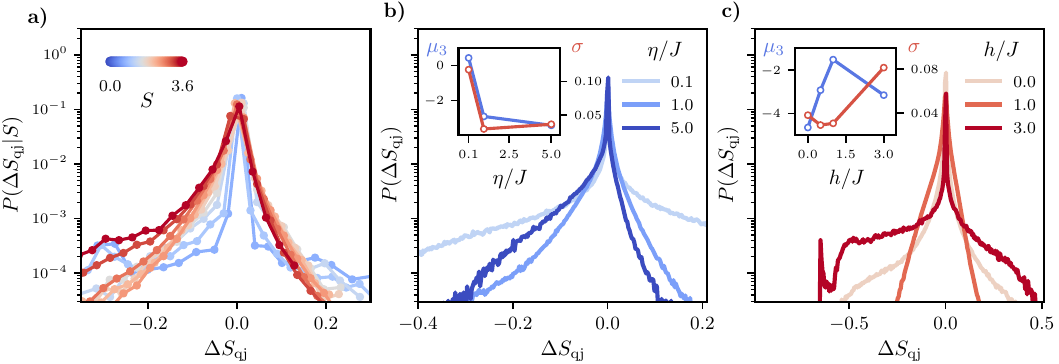}
\caption{ Quantum Jumps Statistics. $a)$ - Distribution of the entanglement change due to quantum jumps $\Delta S_\mathrm{qj}$ conditioned to the entanglement content $S$ for parameters $\gamma =0.5J$, $\eta = J$,  and $h =0.5J$ . $b)$ - Distribution of the entanglement change due to quantum jumps $\Delta S_\mathrm{qj}$ for different pairing parameters $\eta= \{0.1J,~J,~5.0J\}$. The standard deviation $\sigma$ and the skewness $\mu_3$ ($3^{rd}$ moment of the distribution representing the asymmetry of the distribution, it is negative if the distribution is dominated by values below the average and positive for the opposite) are plotted in the inset as a function of $\eta$. The left y-axis is the skewness scale (plotted in blue), and the right y-axis is the standard deviation scale (plotted in red). Other parameters: $h=0.5J$, $\gamma=0.5J$. $c)$ - Distribution of the entanglement change due to quantum jumps $\Delta S_\mathrm{qj}$ for different on-site energy $h= \{0.0J,~0.5J,~1.0J,~3.0J\}$. The standard deviation and the skewness of the distribution $\sigma$ are plotted in the inset as a function of $h/J$ and with the same convention as in $b)$. Other parameters $\gamma=0.5J$, $\eta=J$ \label{fig:entanglement_stats}
}
\end{figure*}

We start presenting the statistics of the entanglement entropy gain and loss for our monitored fermionic chain, focusing on the quantum jump contribution. We plot the histogram in Fig.~\ref{fig:entanglement_stats} $a)$ which shows two main features: (i) the highest probability is for $\Delta S_\mathrm{qj}=0$, meaning that most QJs do not create any change in the entanglement of the state despite that here each QJ removes a particle from the system; (ii) the distribution is asymmetric, reflecting that QJs are more likely to decrease the entanglement. The conditional QJs distribution is also consistent with what has been observed in~\cite{gal2024entanglementdynamicsmonitoredsystems}, where QJs tend to significantly reduce the entanglement when the entanglement content is high, which testifies about the fragility of these highly entangled fermionic states toward quantum jumps. 

In Fig.~\ref{fig:entanglement_stats} $b)$, we study the impact of $\eta/J$ on the distribution of the entanglement change due to QJs. We see that changing $\eta/J$ mainly affects the tails of the distribution, while the statistics around the typical value of $\Delta S_\mathrm{qj}$ is only weakly $\eta$-dependent. For small $\eta/J$, the distribution is broad and slightly asymmetric: quantum jumps that significantly change the entanglement entropy acquire a non-negligible probability. This is consistent with the fact that, as we have seen in Sec.~\ref{sec:U_1_waiting_times}, for $\eta/J=0$ quantum jumps are strongly relevant, depleting the system and driving the entanglement entropy to zero at long times. Upon increasing $\eta/J$, the histogram becomes less broad and more asymmetric, with a higher probability of negative changes to the entanglement entropy, an effect that further increases with $\eta/J$.
To highlight this aspect, we plot in the inset of Fig.~\ref{fig:entanglement_stats} $b)$ the asymmetry of the distribution characterized by its skewness $\mu_3 = \int_{\Delta S_\mathrm{qj}} P(\Delta S_\mathrm{qj}) \left[(\Delta S_\mathrm{qj} - \overline{\Delta S_\mathrm{qj}})^3/\sigma(\Delta S_\mathrm{qj} )^3 \right]$.
The latter, significantly negative at large $\eta/J$ (i.e. jumps tend to decrease the entanglement as reported before), tends to zero when pairing terms are small. On its side, the standard deviation $\sigma(\Delta S_\mathrm{qj})$ is increasing at small $\eta/J$ which confirms the mentioned broadening of the distribution.

In panel $c)$ of Fig.~\ref{fig:entanglement_stats}, we repeat the same analysis on the statistics of entanglement entropy changes due to quantum jumps for different values of the onsite energy $h$.
We see that for $h/J$ going to zero, the distribution is narrow and centered around its typical value, with small asymmetric tails. This explains our earlier observation that for $h=0$ the entanglement dynamics in the monitored case is in excellent agreement with the no-click limit, as shown in the inset of Fig.~\ref{fig:Ising_line} $a)$. Quantum jumps in this regime are present but do not substantially affect the entanglement entropy. We can say that they are an irrelevant perturbation to the no-click limit. This remains true as $h$ increases to values $h/J\simeq 1$, above which we see that the distribution becomes very broad and asymmetric.
This is clear for $h/J=3$, where we see that the statistical weight of quantum jumps with $\Delta S_\mathrm{qj}\simeq 0$ decreases in favor of events that change (decrease) the entanglement entropy by a term of order one. This is the regime where quantum jumps are more relevant, and indeed stronger deviations from the no-click limit were observed in Sec.~\ref{sec:Z2} for large $h/J$.

As anticipated, in addition to their own interest, the entanglement gain and loss distribution allows us to construct a classical model that describes the average entanglement $\overline{S}(t,\ell)$ for a particular size bipartition $\ell$, under the assumption that QJs are Poisson distributed over time with a characteristic average waiting time $\bar{\tau}$, a fact that we have numerically verified. Specifically, we model the entanglement evolution as a random walk with random drift and partial resetting~\cite{evans2011diffusion,Evans_2020,turkeshi2022entanglementtransitionsfrom}. First, we draw randomly the QJs times from a Poisson law with the same average $\bar{\tau}$ as the exact simulation. Once the times of quantum jumps are known, we proceed with the random walk: during the non-Hermitian evolution in between QJs, for a time interval $\tau$, we model the change of entanglement entropy due to the non-Hermitian evolution by $\Delta S_{\rm nH} = \tau ~\delta S_{\rm nH}$,  where $\delta S_{\rm nH}$ is picked with a probability $P(\delta S_{\rm nH}\vert S)$ from the previously computed conditional distribution of the effective non-Hermitian slopes. Once the instantaneous QJ happens, we model the change it induces on the entanglement entropy by $\Delta S_{\rm qj}$, which is drawn with a probability $P(\Delta S_{\rm qj}\vert S)$ from the corresponding probability distribution.  We can formalize the stochastic process above in terms of a classical master equation~(\ref{eqn:master_eq}) for the probability of having an entanglement entropy $S$ at time $t$, that we note  $\mathcal{P}_t(S)$:
\begin{equation}
    \begin{split}
        &\mathcal{P}_{t+dt}(S)=rdt \int_{\Delta S_{\rm qj}} P(\Delta S_{\rm qj}\vert S-\Delta S_{\rm qj}) \mathcal{P}_t(S-\Delta S_{\rm qj})+\\
&+(1-rdt)\int_{\delta S_{\rm nH}} P(\delta S_{\rm nH}\vert S-\delta S_{\rm nH}dt) \mathcal{P}_t(S-\delta S_{\rm nH}dt)
    \end{split}
    \label{eqn:master_eq}
\end{equation}
where the first term describes the jump, which adds a random contribution $\Delta S_{\rm qj}$ with probability $rdt\times P(\Delta S_{\rm qj}\vert S-\Delta S_{\rm qj})$, with the resetting rate $r=1/\bar{\tau}$, while the second one describes the non-Hermitian dynamics which increase the entanglement of a random slope $\delta S_{\rm nH}$ with probability $(1-rdt)\times P(\delta S_{\rm nH}\vert S-\delta S_{\rm nH}dt)$.

Eq.~(\ref{eqn:master_eq}) describes a random walk for the  entanglement entropy, alternating between changes due to jumps and non-Hermitian evolution. These changes are drawn from the entanglement gain and loss distribution discussed above. From this master equation one can obtain a closed form equation for the time evolution for the average entanglement entropy $\overline{S}$ which reads~\cite{gal2024entanglementdynamicsmonitoredsystems}
\begin{equation}
    \frac{d\overline{S}}{dt} = \overline{\delta S_\mathrm{nH}}(S,\ell) +  \overline{\Delta S_\mathrm{qj}}(S,\ell) /  \overline{\tau}.
\end{equation}
This is a balance equation between the average change in entanglement due to the non-Hermitian evolution, $\overline{\delta S_\mathrm{nH}}(S,\ell)$, and the average change in entanglement due to QJs, $\overline{\Delta S_\mathrm{qj}}(S,\ell)$, renormalized by their average frequency, or equivalently the inverse of the average waiting time $\overline{\tau}$. 
The steady-state is then given by the fixed point of this balance equation. For further details, we encourage the reader to refer to~\cite{gal2024entanglementdynamicsmonitoredsystems}. 
\begin{figure}[t]
\begin{centering}
\includegraphics{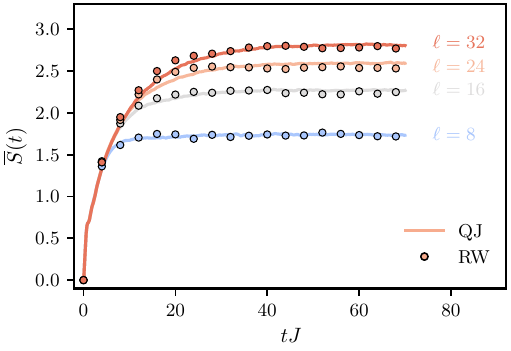}
\end{centering}
\caption{Comparison between the average entanglement obtained from the exact monitored dynamics (QJ) and the phenomenological stochastic model (RW). Parameters $\gamma =0.5J$, $\eta = J$,  and $h =0.5J$, and the conditional distributions can be seen in Fig.~\ref{fig:entanglement_stats} $a)$.  \label{fig:class_mod}}
\end{figure}

In the context of this work, we verify the precision of the classical stochastic model for the average entanglement entropy.  This phenomenological model is designated by RW and the result is presented in Fig.~\ref{fig:class_mod}, where the classical evolution (RW) reproduces well the averaged entanglement evolution (QJ) obtained from the exact evolution of quantum trajectories.

\section{Discussion}\label{sec:discussion}

The results discussed in previous sections demonstrate that the interplay between local particle losses and coherent drive in a non-interacting fermionic chain can lead to a steady-states with a non-trivial entanglement structure, which supports entanglement transitions. In this respect, it is interesting to remark on the relation of our results and previous works.

The role of losses in the entanglement dynamics of monitored fermions was discussed in~\cite{starchl2024generalizedzenoeffectentanglement}. A key difference from our setting is the role of symmetry in the coherent dynamics. In~\cite{starchl2024generalizedzenoeffectentanglement}, the drive was included as an incoherent dissipative process, ultimately leading to a short-ranged entangled state with area-law scaling. In contrast, in our case, the $\mathbb{Z}_2$ symmetry gives rise to a non-trivial non-Hermitian Hamiltonian, which plays a key role in our transition.

In another context, for bosonic particles, monitoring with local linear jump operators has been shown to give rise to a steady-state entanglement with area-law scaling~\cite{minoguchi2022continuousgaussianmeasurements,young2024diffusiveentanglementgrowthmonitored,yokomizo2024measurementinducedphasetransitionfree}, ultimately due to the unbounded bosonic Hilbert space slowing down the entanglement growth~\cite{zhou2021nonunitary}. However, in this case, the monitoring protocol differs from our quantum jumps, as it involves homodyne quantum trajectories, i.e. quantum state diffusion; therefore, a direct comparison is less informative. Moreover, the quantum jump monitoring with bosonic linear quantum jumps breaks the gaussianity of the bosonic state~\cite{Landi2024_prx_quantum}, thus making the problem less suitable for numerical studies.

It is also interesting to comment on the differences and similarities between our setup and the problem of free fermions with pairing and monitoring of particle density, which has been studied extensively~\cite{turkeshi2021measurementinducedentanglement,paviglianiti2023multipartite,gal2024entanglementdynamicsmonitoredsystems}. In this case, the non-Hermitian Hamiltonian retains the same form, with the only difference being in the structure of the jump operator: monitoring the particle density projects locally on a filled site, while monitoring associated with losses projects to an empty site. This leads to significant differences in the average steady-state. In the case of Lindblad dynamics with dephasing, the result is a trivial infinite temperature state, while in the presence of losses, the steady-state can exhibit non-trivial correlations. At the level of entanglement in the quantum trajectories, our results show that the two problems share many of the qualitative features in the phase diagram, which again suggests that a major role in the problem is played by the non-Hermitian Hamiltonian. However, monitoring with the particle density generically leads to a higher value of the entanglement entropy (see Appendix~\ref{appendix_comparison_jumps} for a comparison), with more extended deviations from the no-click limit. 

\section{Conclusions}
\label{sec:conclusion}

In this work, we have studied measurement-induced entanglement transitions in a problem of free-fermions with pairing and particle losses. Specifically, we studied the dynamics of the entanglement entropy in the QJ protocol for continuous monitoring. We have shown that in the presence of U(1) symmetry in the Hamiltonian driving the coherent part of the dynamics, the long-time steady-state is the vacuum for all $\gamma$ (monitoring rate), and so, it has vanishing entanglement. We have used a generalized quasiparticle picture to capture the entanglement entropy dynamics in this regime and computed analytically the waiting time-distribution of quantum jumps, which play a key role for the steady-state of the problem.

Adding a pairing term allows to stabilize a finite density steady-state with non-trivial entanglement structure. In particular, we have demonstrated an entanglement transition driven by the monitoring rate of losses, or equivalently by a coherent coupling such as the onsite energy $h$, between a sub-volume phase with logarithmic scaling of the entanglement entropy and an area-law phase. Furthermore, we have shown that the transition can also be driven (and it is ultimately due to) by the pairing term in the Hamiltonian. Overall, the qualitative picture of this entanglement transition is in close relation with the no-click limit, to which we have compared our findings. We have shown that throughout the phase diagram there are parameter regimes where quantum jumps are relevant (such as for $\eta/J\rightarrow0$) and others where they are mainly irrelevant. To better understand the role of quantum jumps, we have turned to the statistics of entanglement gain and loss, a recently introduced metric that looks at how effective quantum jumps are in changing the entanglement entropy~\cite{gal2024entanglementdynamicsmonitoredsystems}. The results of this analysis highlight that major deviations from the no-click limit arise when the statistics of gain/loss becomes broad and asymmetric, making atypical quantum jumps that induce sizable changes to the entanglement entropy increasingly likely. 

We can identify several extensions of this work that would be worth pursuing in the future.
First, it would be interesting to add interactions in the Hamiltonian and study the fate of the measurement-induced transition, an avenue that is still largely unexplored beyond the case of random circuits. In particular the role of different monitoring processes, such as particle losses, in combination with interactions deserves further work. In this case, the importance of the associated interacting no-click problem remains to be understood. One can foresee that the symmetry's influence in the coherent dynamics (i.e. U(1) vs $\mathbb{Z}_2$) also plays a role in this case, with a U(1)-preserving interacting coherent dynamics leading to the vacuum under losses and a pairing term breaking this symmetry leading to a non-trivial entanglement structure, possibly with a volume-law phase. In Lindbladian problems, i.e., for the average state, it is known that anisotropic coherent couplings give rise to exotic symmetry-breaking patterns and dissipative phase transitions~\cite{lee2013unconventional}. 

Another direction worth exploring is the role of correlated losses for measurement-induced transitions, such as two-body losses. This also breaks the gaussianity of the state, leading to an interacting non-Hermitian Hamiltonian. Once more, this is studied in the Lindbladian case, where these types of dissipative process are known to result in slow dynamics~\cite{Garcia-Ripoll_2009,rossini2021strong,mazza2023dissipative,gerbino2024large}. It would be interesting to investigate their impact on the entanglement of quantum trajectories.

\begin{acknowledgments}
R.D.S acknowledges funding from Erasmus$+$ (Erasmus Mundus program of the European Union) and from the Institute Quantum-Saclay under the project \emph{QuanTEdu-France}. We acknowledge the computational resources on the Coll\`{e}ge de France IPH cluster.
\end{acknowledgments}

\appendix

\begin{figure*}[t!]
\centering
\includegraphics{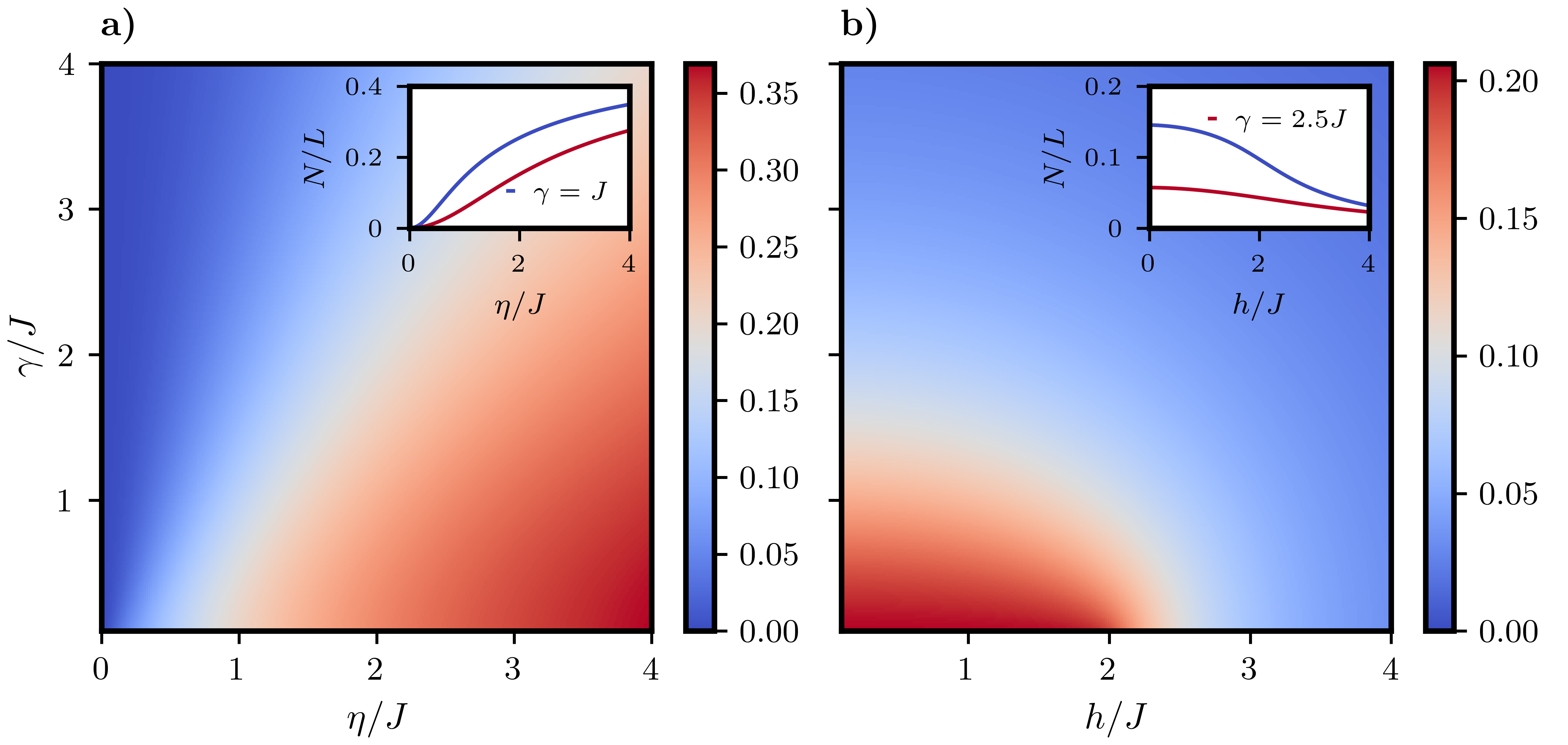}
\caption{Steady-state particle density of the unconditional Lindblad dynamics, for distinct values of the system parameters. In $a)$ $h=0$ and we tune $\gamma/J$ and $\eta/J$, while in $b)$ we fix $\eta=J$ and tne $\gamma/J$ and $h/J$. The inset of both graphs respects the same color scheme: the blue curve corresponds to $\gamma=J$, while the red curve corresponds to $\gamma=2.5J$. The rest of the parameters are the same as stated before.}
\label{fig:particle_density}
\end{figure*}

\section{Dynamics of the Average State and Lindbladian}
\label{sec:lindblad}

In this Appendix, we discuss the dynamics of the average density matrix, which evolves under the Lindblad master equation
\begin{align}
\frac{d\rho}{dt}=-i[\mathcal{H},\rho]+2\gamma\sum_n \left( c_n\rho c_n^{\dagger}-\frac{1}{2}\left\{c_n^{\dagger}c_n,\rho\right\}\right),
\end{align}
with 
$$
\mathcal{H}=-\sum_{n=0}^{L-2}\left[J \;c_{n}^{\dagger}c_{n+1}+\eta \;c_{n}^{\dagger}c_{n+1}^{\dagger}+\text{h.c}\right]+h\sum_{n=0}^{L-1}c_{n}^{\dagger}c_{n}\,,
$$ 
as given in the main text. The calculation of expectation values of operators using the Quantum trajectories and Linbladian must be the same for linear observables in the density matrix and we have verified this is indeed the case, to confirm the convergence of the Monte Carlo wave function sampling.
For the model above, assuming periodic boundary conditions, it is possible to determine in close form the steady-state solution for the 2-point correlation matrix, since the Hamiltonian is quadratic and the jump operators are linear in the fermionic fields. This is done by first writing the equations of motion for a generic observable $\mathcal{O}$,
\begin{equation}
i\dfrac{d}{dt}\left\langle \mathcal{O}\right\rangle =\left\langle \mathcal{O}\mathcal{H}_{\text{eff}}-\mathcal{H}_{\text{eff}}^{\dagger}\mathcal{O}\right\rangle +2\gamma i\sum_{k}\left\langle c_{k}^{\dagger}\mathcal{O}c_{k}\right\rangle,
\end{equation}
where $\mathcal{H}_{\text{eff}}$ is the same effective non-Hermitian Hamiltonian defined in the main text (Eq.~\ref{eqn:Heff2}), which in momentum space reads
\begin{equation}
\mathcal{H}_{\text{eff}}=\sum_k \dfrac{1}{2}\Psi_{k}^{\dagger}\begin{pmatrix}\varepsilon_{k}-i\gamma/2 & -\Delta_{k}\\
-\Delta_{k}^{\ast} & -\varepsilon_{-k}+i\gamma/2
\end{pmatrix}\Psi_{k},
\end{equation}
where $\varepsilon_k = h-2J\cos(k)$, $\Delta_{k}=2i\eta\sin(ka)$ and $\Psi^\dagger_k =\begin{pmatrix}c_{k}^{\dagger} & c_{-k}\end{pmatrix}$. To fully specify the system state, we need to compute the equation of motion for the following two by two correlation matrix,
\begin{equation}
\mathcal{C}_{k}=\begin{pmatrix}\left\langle c_{k}^{\dagger}c_{k}\right\rangle  & \left\langle c_{k}^{\dagger}c_{-k}^{\dagger}\right\rangle \\
\left\langle c_{-k}c_{k}\right\rangle  & \left\langle c_{-k}c_{-k}^{\dagger}\right\rangle 
\end{pmatrix}.
\end{equation}

After simple manipulations, the equations of motion take the form
\begin{widetext}
\begin{equation}
\begin{aligned}
i\partial_{t}\left\langle c_{k}^{\dagger}c_{k}\right\rangle  & =-\Delta_{k}\left(\left\langle c_{k}^{\dagger}c_{-k}^{\dagger}\right\rangle +\left\langle c_{-k}c_{k}\right\rangle \right)-2i\gamma\left\langle c_{k}^{\dagger}c_{k}\right\rangle, \\
i\partial_{t}\left\langle c_{k}^{\dagger}c_{-k}^{\dagger}\right\rangle  & =-2\varepsilon_{k}\left\langle c_{k}^{\dagger}c_{-k}^{\dagger}\right\rangle +\Delta_{k}\left(\left\langle c_{k}^{\dagger}c_{k}\right\rangle -\left\langle c_{-k}c_{-k}^{\dagger}\right\rangle \right)-2i\gamma\left\langle c_{k}^{\dagger}c_{-k}^{\dagger}\right\rangle, \\
i\partial_{t}\left\langle c_{-k}c_{k}\right\rangle  & =2\varepsilon_{k}\left\langle c_{-k}c_{k}\right\rangle +\Delta_{k}\left(\left\langle c_{k}^{\dagger}c_{k}\right\rangle -\left\langle c_{-k}c_{-k}^{\dagger}\right\rangle \right)-2i\gamma\left\langle c_{-k}c_{k}\right\rangle. 
\end{aligned}
\end{equation}
\end{widetext}
For finite $\gamma$, we can solve for the steady-state solutions,
\begin{equation}
\begin{aligned}
\left\langle c_{k}^{\dagger}c_{k}\right\rangle _{\text{ss}}&=\dfrac{1}{2}\dfrac{\Delta_{k}^{2}}{\left(\Delta_{k}^{2}-\varepsilon_{k}^{2}-\gamma^{2}\right)},\\
\left\langle c_{k}^{\dagger}c_{-k}^{\dagger}\right\rangle _{\text{ss}} &= \dfrac{1}{2}\dfrac{\Delta_{k}\left(\varepsilon_{k}-i\gamma\right)}{\left(\Delta_{k}^{2}-\varepsilon_{k}^{2}-\gamma^{2}\right)}.
\end{aligned}
\end{equation}
The total particle number is obtained by summing over all momenta in the First Brillouin Zone, which, in the thermodynamic limit, is equivalent to

\begin{align}
\left\langle N\right\rangle  & =\dfrac{L}{2}\int_{-\pi}^{\pi}\dfrac{dk}{2\pi}\dfrac{\Delta_{k}^{2}}{\left(\Delta_{k}^{2}-\varepsilon_{k}^{2}-\gamma^{2}\right)}.
\end{align}
In Fig.~\ref{fig:particle_density} the steady-state density of the system is given as a function of the system parameters. In general, this integral does not have a closed analytical solution. Nevertheless, there are some limiting cases in which it is exactly solvable. For example, when $h=0$ and $\eta=J$,
\begin{equation}
\left\langle N\right\rangle = L\dfrac{J^{2}}{4J^{2}+\gamma^{2}}.
\end{equation}
In the limit $\gamma\gg J$, the particle density is decreasing with $\gamma^{-2}$. On the other hand, without the pairing term, the steady-state corresponds to the fermionic vacuum, since in this case, all steady-state 2-point correlation functions are zero. In this scenario, the equations of motion indicate that
\begin{equation}
    \expected{c^\dagger_k c_k \left(t \right)} = \expected{c^\dagger_k c_k \left(0 \right)}e^{-2\gamma t}.
\end{equation}
This corresponds to the quasiparticle density for $\eta = 0$, since, in this case, the Hamiltonian is diagonal in the operators $c^\dagger_k c_k$.

Moreover, in this case it follows directly that the the total particle number decreases also exponentially over time, according to
\begin{equation}
\left\langle N\right\rangle =\left\langle N\right\rangle _{0}e^{-2\gamma t}.
\end{equation}
Overall we conclude that the steady-state of the system under the Lindblad dynamics does not show any signature of phase transition, in accordance with general results~\cite{zhang2022criticality} which predict a finite correlation length. We also note that this remains the case in presence of partial measurement averaging~\cite{paviglianiti2024breakdownmeasurementinducedphasetransitions}.

\begin{figure}[t]
\centering
\includegraphics{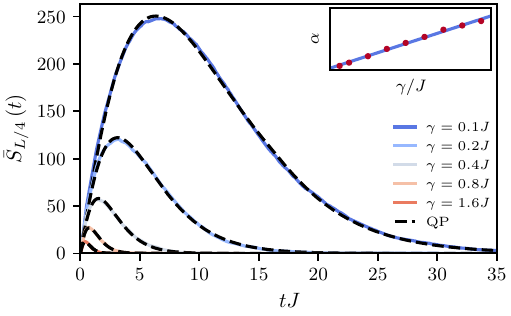}
\caption{Time evolution of the entanglement entropy for different values of the monitoring rate. The coefficient $\alpha$ corresponding to the effective decay of the quasiparticle density is given by $\alpha = a \cdot \gamma + b$, where $a=\left(2.36 \pm 0.06 \right)$ and $b=\left( 0.06 \pm 0.06 \right)$. Other parameters: $\eta=J$, $h=0$ and $L=128$.} 
\label{fig:quasi_particle_fit}
\end{figure} 

\section{The Effective Quasiparticle Picture}
\label{app:quasiparticle}

In this Appendix, we provide additional numerical evidence to support the quasiparticle picture written in the main text. Specifically, in Fig.~\ref{fig:quasi_particle_fit}, the time evolution of the entanglement entropy is presented for a chain initially set at half-filling with total length L=$128$ and different values of the monitoring rate $\gamma$. To reproduce these data with the quasiparticle picture in the main text, we use the decay rate $\alpha$ of the quasiparticle occupation number $\langle n_k(t)\rangle\sim e^{-\alpha t}$ as a fitting parameter. In the inset, we show that, as expected $\alpha\propto \gamma$, although the prefactor is renormalized with respect to the value obtained within the Lindblad evolution.

\begin{figure*}[t!]
\centering
\includegraphics{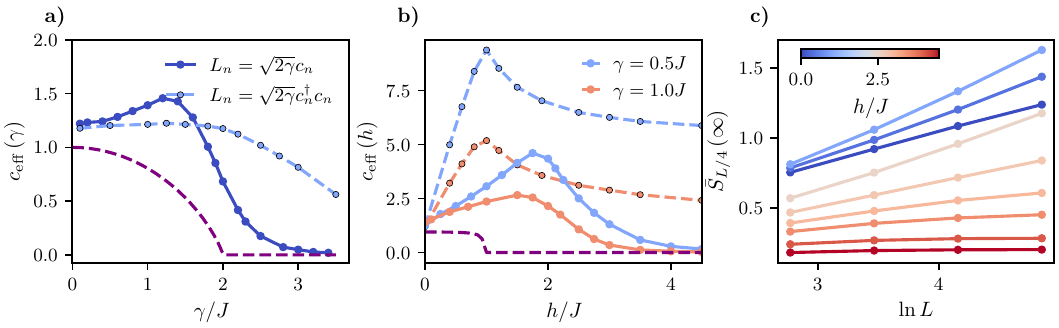}
\caption{$a)$ - Effective central charge as a function of the measurement rate $\gamma/J$. The solid line corresponds to the monitoring with the linear quantum jumps ($L_n = \sqrt{2\gamma} c_n$) and the dashed line corresponds to the monitoring of density ($L_n = \sqrt{2\gamma} c_n^\dagger c_n$). The purple dashed line shows the no-click limit. Other parameters: $\eta=J$ and $h=0$. $b)$ - Effective central charge as a function of the onsite potential for different values of $\gamma/J$. Similarly, the solid lines correspond to linear quantum jump monitoring, while the dashed lines correspond to density monitoring. The purple dashed line shows the no-click limit for $\gamma=0.5J$. Other parameters: $\eta=J$. $c)$ - Steady state average entanglement entropy as a function of the total size of the system for different strengths of the local potential $h/J$ with the montoring rate $\gamma=J$. Other parameters: $\eta=J$. \label{fig:comp_dens_lin}}
\end{figure*}
\section{Comparison with the density monitoring}
\label{appendix_comparison_jumps}
As we mentioned in the main text, it is particular interesting to compare our results for the model of free fermions with pairing and losses with those obtained, for the same model, when monitoring the particle density~\cite{turkeshi2021measurementinducedentanglement,turkeshi2022entanglementtransitionsfrom,gal2024entanglementdynamicsmonitoredsystems}. Indeed, the non-Hermitian Hamiltonian takes the same form in the two problems, given again by Eq.~(\ref{eqn:Heff}) of the main text, and the only difference is in the structure of the jump operators. Therefore, the differences in the phase diagrams directly reflect the action of the quantum jumps. Previous numerical results~\cite{gal2024entanglementdynamicsmonitoredsystems} have shown that density monitoring leads to a qualitatively similar phase diagram, with a sub-volume logarithmic phase for the entanglement entropy at weak monitoring and an area-law phase at large monitoring.

Here we compare the dependence of the effective central charge as a function of $\gamma$ (Fig.\ref{fig:comp_dens_lin} $a)$) and $h$ (Fig.\ref{fig:comp_dens_lin} $b)$) for two different jump operators: $L_n = \sqrt{2\gamma} c_n$ (particle loss) and $L_n = \sqrt{2\gamma} c_n^\dagger c_n$ (density measurement). For completeness, in Fig.\ref{fig:comp_dens_lin} $c)$ we show the data for the entanglement entropy from which the effective central charge is extracted in panel b) for $\gamma=J$ and particle losses monitoring.
As we see from panel a) at a low monitoring rate, the effective central charges obtained with both quantum jump operators tend to converge to similar values, as the effect of QJs is reduced and the non-Hermitian Hamiltonian becomes the primary factor driving the evolution. The dependence on the onsite potential is similar in both cases, showing an enhancement of the logarithmic phase for small and intermediate ratios of $h/J$. An increase in $h/J$ results in a reduction in the effective central charge. This reduction is more pronounced and occurs at lower values of the onsite potential when monitoring with losses.

\section{Details on the Numeric Implementation}
\label{sec:details}
The numerical simulations performed in this work were performed by combining the high-order Monte Carlo Wave Function algorithm ~\cite{Daley2014_quantum_trajec_review,Landi2024_prx_quantum} with the Faber Polynomial algorithm present in~\cite{soares2024nonunitaryquantummanybodydynamics}. Specifically, in between quantum jumps, the non-unitary propagation of the state is obtained by expanding the time evolution operator in a Faber polynomial series,
\begin{equation}
\begin{aligned}
e^{-i\mathcal{H}_{\rm eff}t}\ket{\Psi \left(t_0\right)}&=\sum_{n=0}^{+\infty}c_n \left(t \right)F_{n}\left(\dfrac{\mathcal{H}_{\rm eff}}{\lambda}\right) \ket{\Psi \left(t_0\right)},\\
c_n \left(t \right)&=e^{-i\lambda\gamma_{0}t}\left(\dfrac{-i}{\sqrt{\gamma_{1}}}\right)^{n}J_{n}\left(2\sqrt{\gamma_{1}}\lambda t\right),
\end{aligned}
\label{eq:Faber_expansion}
\end{equation}
where $F_n$ is the $\rm n^{th}$ Faber Polynomial, $J_n$ is the $\rm n^{th}$ Bessel function of the first kind, $\lambda$ is a rescaling parameter and $\gamma_0$ and $\gamma_1$ are related with the bounds of the spectrum of the non-Hermitian Hamiltonian, consult~\cite{soares2024nonunitaryquantummanybodydynamics} for the specific details. In practice, the series is truncated up to a finite order, and the action of $F_n \left(\mathcal{H}_{\rm eff}/\lambda \right) \ket{\Psi \left(t_0 \right)}$ is done efficiently by using the recurrence relation satisfied by the Faber Polynomials, namely,
\begin{equation}
\begin{aligned}
\ket{\Psi_{0}} & =\ket{\Psi\left(t_{0}\right)}\\
\ket{\Psi_{1}} & =\left(\tilde{\mathcal{H}}_{\rm eff}-\gamma_{0}\right)\ket{\Psi_{0}}\\
\ket{\Psi_{2}} & =\left(\tilde{\mathcal{H}}_{\rm eff}-\gamma_{0}\right)\ket{\Psi_{1}}-2\gamma_{1}\ket{\Psi_{0}}\\
\ket{\Psi_{n+1}} & =\left(\tilde{\mathcal{H}}_{\rm eff}-\gamma_{0}\right)\ket{\Psi_{n}}-\gamma_{1}\ket{\Psi_{n-1}},\quad n > 2,
\end{aligned}
\label{eq:req_faber}
\end{equation}
where $\ket{\Psi_n} = F_n \left(\mathcal{H}_{\rm eff}/\lambda \right) \ket{\Psi\left(t_0\right)}$. This approach is highly memory efficient, as it requires storing only the last two vectors in memory, $\ket{\Psi_{n-1}}$ and $\ket{\Psi_n}$, to compute the next one, $\ket{\Psi_{n+1}}$. After each time step, the state is normalized. In practice, the expansion in Eq.~\eqref{eq:Faber_expansion} is done at the level of the single-particle Hamiltonian (represented on the Nambu basis) by using the Gaussian nature of the state to our advantage. The state is parameterized in the usual way,
\begin{equation}
\ket{\Psi(t)}=\mathcal{N}\left(t\right)\exp\left(-\dfrac{1}{2}\sum_{m,n}\left[\left(U_{t}^{\dagger}\right)^{-1}V_{t}^{\dagger}\right]_{m,n}c_{m}^{\dagger}c_{n}^{\dagger}\right)\ket{\text{vac}},
\end{equation}
where $\mathcal{N}$ enforces the correct normalization and $U$ and $V$ are $\text{L} \times{L}$ matrices. These matrices evolve according to,
\begin{equation}
    \begin{pmatrix}U_{t}\\
    V_{t}
\end{pmatrix}=e^{-2i\mathbb{H}_{\text{eff}}}\begin{pmatrix}U\left(0\right)\\
V\left(0\right)
\end{pmatrix},
\end{equation}
where $\mathbb{H}_{\rm eff}$ is the single-particle non-Hermitian Hamiltonian in the Nambu representation. For further details, see~\cite{mbeng2024thequantum,gal2024entanglementdynamicsmonitoredsystems}.

Following a quantum jump, the state of the system is modified at the level of the correlation matrix. This modification is performed using standard procedures~\cite{turkeshi2022entanglementtransitionsfrom,gal2024entanglementdynamicsmonitoredsystems}, where to perform the update one relies on the Gaussianity of the state before the quantum jump. For instance, if the quantum jump occurs at site, $\ell$, the post-jump state is simply given as
\begin{equation}
\ket{\tilde{\Psi}}=\dfrac{c_{\ell}\ket{\Psi}}{\bra{\Psi}c^\dagger_{\ell}c_{\ell}\ket{\Psi}}.
\end{equation}
The post-jump regular and anomalous 2-point correlation functions are given by the following 4-point correlation functions evaluated in the pre-jump state,
\begin{equation}
\begin{aligned}
G_{m,n}^{J} & =\dfrac{\left\langle c_{\ell}^{\dagger}c_{m}^{\dagger}c_{n}c_{\ell}\right\rangle}{\left\langle c^\dagger_\ell c_\ell \right\rangle},\; F_{m,n}^{J} & =\dfrac{\left\langle c_{\ell}^{\dagger}c_{m}^{\dagger}c_{n}^{\dagger}c_{\ell}\right\rangle}{\left\langle c^\dagger_\ell c_\ell \right\rangle}.
\end{aligned}
\end{equation}
The denominator can be expanded using Wick's theorem,
\begin{equation}
\begin{aligned}
G_{m,n}^{J} &=\left\langle c_{\ell}^{\dagger}c_{\ell}\right\rangle \left\langle c_{m}^{\dagger}c_{n}\right\rangle -\left\langle c_{\ell}^{\dagger}c_{n}\right\rangle \left\langle c_{m}^{\dagger}c_{\ell}\right\rangle+ \\
&+\left\langle c_{\ell}^{\dagger}c_{m}^{\dagger}\right\rangle \left\langle c_{n}c_{\ell}\right\rangle,\\
F_{m,n}^{J} & =\left\langle c_{\ell}^{\dagger}c_{\ell}\right\rangle \left\langle c_{m}^{\dagger}c_{n}^{\dagger}\right\rangle +\left\langle c_{\ell}^{\dagger}c_{m}^{\dagger}\right\rangle \left\langle c_{n}^{\dagger}c_{\ell}\right\rangle +\\
&-\left\langle c_{\ell}^{\dagger}c_{n}^{\dagger}\right\rangle \left\langle c_{m}^{\dagger}c_{\ell}\right\rangle.
\end{aligned}
\end{equation}

This results in the following update rule after a quantum jump:
\begin{equation}
G_{mn}^{J}=\begin{cases}
0, & m=n=\ell,\\
0, & m=\ell\;\text{or}\;n=\ell,\\
G_{mn}-\dfrac{G_{\ell n}G_{m\ell}-F_{\ell m}\left(F^{\ast}\right)_{n,\ell}}{G_{\ell\ell}}, & \text{otherwise}.
\end{cases}
\end{equation}
For the anomalous correlation function,
\begin{equation}
F_{mn}^{J}=\begin{cases}
0, & m=n=\ell,\\
0, & m=\ell\;\text{and}\;n\neq\ell,\\
0 & m\neq\ell\;\text{and}\; n=\ell,\\
F_{mn}+\dfrac{F_{\ell m}G_{n\ell}-F_{\ell n}G_{m\ell}}{G_{\ell,\ell}}, & \text{otherwise}.
\end{cases}
\end{equation}

\bibliography{ref_measurements} 

\end{document}